\RequirePackage{lineno}
\documentclass[preprint, 12pt]{aastex}
\begin{document}
\linenumbers

\title{VERITAS Observations of the TeV Binary LS I +61$^{\circ}$ 303 During 2008-2010}
\author{
V.~A.~Acciari\altaffilmark{1},
E.~Aliu\altaffilmark{2},
T.~Arlen\altaffilmark{3},
T.~Aune\altaffilmark{4},
M.~Beilicke\altaffilmark{5},
W.~Benbow\altaffilmark{1},
S.~M.~Bradbury\altaffilmark{6},
J.~H.~Buckley\altaffilmark{5},
V.~Bugaev\altaffilmark{5},
K.~Byrum\altaffilmark{7},
A.~Cannon\altaffilmark{8},
A.~Cesarini\altaffilmark{9},
L.~Ciupik\altaffilmark{10},
E.~Collins-Hughes\altaffilmark{8},
M.~P.~Connolly\altaffilmark{9},
W.~Cui\altaffilmark{11},
R.~Dickherber\altaffilmark{5},
C.~Duke\altaffilmark{12},
M.~Errando\altaffilmark{2},
A.~Falcone\altaffilmark{13},
J.~P.~Finley\altaffilmark{11},
G.~Finnegan\altaffilmark{14},
L.~Fortson\altaffilmark{15},
A.~Furniss\altaffilmark{4},
N.~Galante\altaffilmark{1},
D.~Gall\altaffilmark{11},
G.~H.~Gillanders\altaffilmark{9},
S.~Godambe\altaffilmark{14},
S.~Griffin\altaffilmark{16},
J.~Grube\altaffilmark{10},
R.~Guenette\altaffilmark{16},
G.~Gyuk\altaffilmark{10},
D.~Hanna\altaffilmark{16},
J.~Holder\altaffilmark{17,*},
G.~Hughes\altaffilmark{18},
C.~M.~Hui\altaffilmark{14},
T.~B.~Humensky\altaffilmark{19},
P.~Kaaret\altaffilmark{20},
N.~Karlsson\altaffilmark{15},
M.~Kertzman\altaffilmark{21},
D.~Kieda\altaffilmark{14},
H.~Krawczynski\altaffilmark{5},
F.~Krennrich\altaffilmark{22},
M.~J.~Lang\altaffilmark{9},
S.~LeBohec\altaffilmark{14},
G.~Maier\altaffilmark{18},
P.~Majumdar\altaffilmark{3},
S.~McArthur\altaffilmark{5},
A.~McCann\altaffilmark{16},
P.~Moriarty\altaffilmark{23},
R.~Mukherjee\altaffilmark{2},
R.~A.~Ong\altaffilmark{3},
M.~Orr\altaffilmark{22},
A.~N.~Otte\altaffilmark{4},
N.~Park\altaffilmark{19},
J.~S.~Perkins\altaffilmark{1},
M.~Pohl\altaffilmark{18,24},
H.~Prokoph\altaffilmark{18},
J.~Quinn\altaffilmark{8},
K.~Ragan\altaffilmark{16},
L.~C.~Reyes\altaffilmark{19},
P.~T.~Reynolds\altaffilmark{25},
E.~Roache\altaffilmark{1},
H.~J.~Rose\altaffilmark{6},
J.~Ruppel\altaffilmark{24},
D.~B.~Saxon\altaffilmark{17},
M.~Schroedter\altaffilmark{22},
G.~H.~Sembroski\altaffilmark{11},
G.~Demet~Senturk\altaffilmark{26},
A.~W.~Smith\altaffilmark{7,27*},
D.~Staszak\altaffilmark{16},
G.~Te\v{s}i\'{c}\altaffilmark{16},
M.~Theiling\altaffilmark{1},
S.~Thibadeau\altaffilmark{5},
K.~Tsurusaki\altaffilmark{20},
A.~Varlotta\altaffilmark{11},
V.~V.~Vassiliev\altaffilmark{3},
S.~Vincent\altaffilmark{14},
M.~Vivier\altaffilmark{17},
S.~P.~Wakely\altaffilmark{19},
J.~E.~Ward\altaffilmark{8},
T.~C.~Weekes\altaffilmark{1},
A.~Weinstein\altaffilmark{3},
T.~Weisgarber\altaffilmark{19},
D.~A.~Williams\altaffilmark{4},
B.~Zitzer\altaffilmark{11}
}
\altaffiltext{*}{Corresponding Authors: awsmith@hep.anl.gov, jholder@physics.udel.edu}
\altaffiltext{1}{Fred Lawrence Whipple Observatory, Harvard-Smithsonian Center for Astrophysics, Amado, AZ 85645, USA}
\altaffiltext{2}{Department of Physics and Astronomy, Barnard College, Columbia University, NY 10027, USA}
\altaffiltext{3}{Department of Physics and Astronomy, University of California, Los Angeles, CA 90095, USA}
\altaffiltext{4}{Santa Cruz Institute for Particle Physics and Department of Physics, University of California, Santa Cruz, CA 95064, USA}
\altaffiltext{5}{Department of Physics, Washington University, St. Louis, MO 63130, USA}
\altaffiltext{6}{School of Physics and Astronomy, University of Leeds, Leeds, LS2 9JT, UK}
\altaffiltext{7}{Argonne National Laboratory, 9700 S. Cass Avenue, Argonne, IL 60439, USA}
\altaffiltext{8}{School of Physics, University College Dublin, Belfield, Dublin 4, Ireland}
\altaffiltext{9}{School of Physics, National University of Ireland Galway, University Road, Galway, Ireland}
\altaffiltext{10}{Astronomy Department, Adler Planetarium and Astronomy Museum, Chicago, IL 60605, USA}
\altaffiltext{11}{Department of Physics, Purdue University, West Lafayette, IN 47907, USA }
\altaffiltext{12}{Department of Physics, Grinnell College, Grinnell, IA 50112-1690, USA}
\altaffiltext{13}{Department of Astronomy and Astrophysics, 525 Davey Lab, Pennsylvania State University, University Park, PA 16802, USA}
\altaffiltext{14}{Department of Physics and Astronomy, University of Utah, Salt Lake City, UT 84112, USA}
\altaffiltext{15}{School of Physics and Astronomy, University of Minnesota, Minneapolis, MN 55455, USA}
\altaffiltext{16}{Physics Department, McGill University, Montreal, QC H3A 2T8, Canada}
\altaffiltext{17}{Department of Physics and Astronomy and the Bartol Research Institute, University of Delaware, Newark, DE 19716, USA}
\altaffiltext{18}{DESY, Platanenallee 6, 15738 Zeuthen, Germany}
\altaffiltext{19}{Enrico Fermi Institute, University of Chicago, Chicago, IL 60637, USA}
\altaffiltext{20}{Department of Physics and Astronomy, University of Iowa, Van Allen Hall, Iowa City, IA 52242, USA}
\altaffiltext{21}{Department of Physics and Astronomy, DePauw University, Greencastle, IN 46135-0037, USA}
\altaffiltext{22}{Department of Physics and Astronomy, Iowa State University, Ames, IA 50011, USA}
\altaffiltext{23}{Department of Life and Physical Sciences, Galway-Mayo Institute of Technology, Dublin Road, Galway, Ireland}
\altaffiltext{24}{Institut f\"ur Physik und Astronomie, Universit\"at Potsdam, 14476 Potsdam-Golm,Germany}
\altaffiltext{25}{Department of Applied Physics and Instrumentation, Cork Institute of Technology, Bishopstown, Cork, Ireland}
\altaffiltext{26}{Columbia Astrophysics Laboratory, Columbia University, New York, NY 10027, USA}
\altaffiltext{27}{Department of Physics and Astronomy, Northwestern University, Evanston, IL, 60208, USA}

\begin{abstract}

We present the results of observations of the TeV binary
LS~I~+61$^{\circ}$~303 with the VERITAS telescope array between 2008
and 2010, at energies above 300 GeV. In the past, both ground-based gamma-ray telescopes VERITAS and MAGIC have reported detections
of TeV emission near the apastron phases of the binary orbit.  The
observations presented here show no strong evidence for TeV emission
during these orbital phases; however, during observations taken in
late 2010, significant emission was detected from the source close to
the phase of superior conjunction (much closer to periastron passage) at a 5.6 standard deviation ( 5.6 $\sigma$) post-trials significance. In total, between October 2008 and December
2010 a total exposure of 64.5 hours was accumulated with VERITAS on
LS~I~+61$^{\circ}$~303, resulting in an excess at the 3.3$\sigma$
significance level for constant emission over the entire integrated
dataset. The flux upper limits derived for emission during the previously reliably active TeV phases (i.e. close to apastron) are less than 5$\%$ of the Crab Nebula flux in the same energy range. This result stands in apparent contrast to previous observations by both MAGIC and VERITAS which detected the source during these phases at $\>$10$\%$ of the Crab Nebula flux.  During the two year span of
observations, a large amount of X-ray data were also accrued on
LS~I~+61$^{\circ}$~303 by the
\textit{Swift} X-ray Telescope (XRT) and the \textit{Rossi X-ray
Timing Explorer} (RXTE) Proportional Counter Array (PCA). We find no evidence for a correlation between emission in the X-ray
and TeV regimes during 20 directly overlapping observations. We also
comment on data obtained contemporaneously by the \textit{Fermi}
Large Area Telescope (LAT).

\end{abstract}

\section{Introduction}

The high-mass X-ray binary LS~I~+61$^{\circ}$~303 is one of only three
such systems firmly identified as a source of TeV gamma rays. Despite
many years of observations across the electromagnetic spectrum, the
system remains, in some respects, poorly characterized. Known to be
the pairing of a massive B0 Ve star and a compact object of unknown
nature (Hutchings and Crampton 1981, Casares 2005),
LS~I~+61$^{\circ}$~303 has been known historically for its energetic
outbursts at radio, X-ray, GeV, and TeV wavelengths (Gregory et al
2002, Greiner and Rau 2001, Harrison et al. 2000, Abdo et al. 2009,
Albert et al. 2008, Acciari et al. 2008), all of these showing
correlation with the 26.5 day orbital cycle of the compact object in
its path around the Be star. According to the most recent radial
velocity measurements, the orbit is elliptical ($e=0.537\pm0.034$),
with periastron passage determined to occur around phase $\phi=0.275$,
apastron passage at $\phi=0.775$, superior conjunction at $\phi=0.081$
and inferior conjunction at $\phi=0.313$ (Aragona et al. 2009). The
non-thermal behavior of the source is well studied, yet poorly
understood. The radio emission presents a well-defined periodicity
with strong flares occuring periodically near apastron passage, along
with an additional modulation on a 4.6 year timescale (Gregory
2002). The detection of extended structures in radio observations
originally identified LS~I~+61$^{\circ}$~303 as a potential
microquasar, with high energy emission produced in jets driven by
accretion onto the compact object, presumably a black hole (Massi et
al 2001). High-resolution VLBA observations indicate that the radio
structures are not persistent, however, and can be more easily
explained by the interaction between a pulsar wind and the wind of the
stellar companion (Dhawan et al. 2006, Albert et al. 2008), although
alternative interpretations are still possible (Romero et al. 2007, 
Massi and Zimmerman 2010). The full range of arguments in favor of
LS~I~+61$^{\circ}$~303 as a non-accreting pulsar system have been
summarized by Torres (2010a).

The X-ray behavior is well studied (see Smith et al. (2009) for a
review), most comprehensively in Torres et al. (2010b), in which the
authors present an analysis of RXTE-PCA observations taken over four
years, covering 35 full orbital cycles. They show that, while the
X-ray emission from the source is indeed periodic, there is
variability of the orbital profiles on multiple timescales, from
individual orbits up to years. Additionally, intense X-ray flares have
been observed (Smith et al. 2009 and Torres et al. 2010b), during which
the flux increases by up to a factor of five and variability on the
timescale of a few seconds is observed. It is noted, however, that, due
to the relatively large RXTE-PCA field-of-view ($\sim1^{\circ}$ FWHM),
the possibility that these flares are due to an unrelated source in
the same field cannot be ruled out.

High-energy (HE; 30~MeV-30~GeV) gamma-ray emission, spatially
coincident with LS~I~+61$^{\circ}$~303, although with large positional
errors, was first detected by COS-B (Hermsen et al. 1977) and,
subsequently, by EGRET (Tavani et al. 1998). The detection of a variable
very high energy (VHE; 30~GeV-30~TeV) gamma-ray source at the location
of LS~I~+61$^{\circ}$~303 with MAGIC (Albert et al. 2006), later
confirmed by VERITAS (Acciari et al. 2008), completed the
identification of this source as a gamma-ray binary.  The TeV emission
reported by both experiments for observations made prior to 2008 is
spread over approximately one quarter of the orbit, with a peak around
apastron (orbital phase $\phi$ = 0.775). \textit{Fermi}-LAT
observations provide the definitive HE detection and have revealed a
number of interesting features (Abdo et al. 2009). The HE emission
reported in the detection paper (based on observations from August
2008 until March 2009) is modulated at the orbital period (peaking
slightly after periastron ($\phi$= 0.225)), although recent results
from the LAT show different behavior (see Summary and
Discussion). The overall Fermi spectrum shows a sharp exponential
cutoff at 6~GeV.

The existence of orbital modulation in the gamma-ray flux is often
explained by the varying efficiency of the inverse-Compton process
around the orbit, although we note that many alternative explanations
exist (see e.g. Sierpowska-Bartosik and Torres (2009) for a
review). In this scenario, inverse-Compton gamma-ray production along
our line of sight is most efficient at superior conjunction ($\phi$=
0.081), where stellar photons interact head-on with energetic leptons
produced either directly in the pulsar wind or in the pulsar
wind/stellar wind shock interaction region. The density of stellar
photons may also play a role in the efficiency of gamma-ray production,
with the highest density occurring at periastron. The TeV flux is
further modulated by photon-photon absorption around the orbit, which
peaks near superior conjunction and may dominate over the modulation
effects due to production efficiency at energies above $\sim$30
GeV. Orbital modulation of the GeV and TeV flux, with large
differences between the lightcurves observed in each energy band, is
therefore not unexpected. Other effects, for example Doppler boosting
of the emission (Dubus et al. 2010), clumps in the wind of the Be star
(Araudo 2009), tidal disruption of the accretion disk (Romero et
al. 2007) or cascading of high-energy photons to lower energies may
also play a role, and provide a better fit of the models to the
observations. However, it should be noted  that the orbital inclination
of the system is poorly constrained, and this complicates the modeling of
emission from this system.

Substantial effort has been invested in constructing multiwavelength
datasets on LS~I~+61$^{\circ}$~303. Preliminary studies including TeV
data (Acciari et al. 2009 and Albert et al. 2008) combined long-term
(multi-orbit) observations at both X-ray and TeV wavelengths. Acciari
et al. (2009) did not detect any correlation between emission at X-ray
and TeV wavelengths with observations taken by VERITAS,
\textit{Swift}-XRT, and RXTE-PCA. However, the multiwavelength
observations used were not strictly overlapping, which is problematic
given the observed fast (several seconds) variability in the X-ray
regime detailed in Smith et al. (2009).  Further X-ray/TeV observations
reported in Anderhub et al. (2009) detect a correlation between X-ray
and TeV emission during a TeV outburst over a single orbital
cycle. While the authors show a correlation (with a 0.5$\%$
probability of being produced from independent datasets), the longer-scale correlation behavior of the source outside of the observed flare
is not well studied.

In this paper we summarize the results of VERITAS observations of
LS~I~+61$^{\circ}$~303 between 2008 and 2010, including a comparison
with strictly contemporaneous X-ray observations. These are the first
reported TeV observations since the launch of \textit{Fermi} in June
2008, and so they provide an opportunity to examine the broadband high-energy (100~MeV - 10~TeV) emission from the source using
contemporaneous observations.

\section{VERITAS Observations}

The VERITAS array of imaging atmospheric Cherenkov telescopes (IACTs),
located in southern Arizona (1250 m.a.s.l., 31$^{\circ}$40' N,
110$^{\circ}$57' W), is composed of four, 12m diameter, 12m focal
length telescopes, each with a Davies-Cotton tessellated mirror
structure of 345 mirror facets (total mirror area of 110 m$^{2}$) (Ong
et al. 2009) . Each telescope focuses Cherenkov light from particle
showers onto a 499-pixel photomultiplier tube (PMT) camera. The pixels
have an angular spacing of 0.15$^{\circ}$, resulting in a camera field of
view of 3.5$^{\circ}$. During the summer months of 2009, one of the
telescopes in the array was relocated, creating a more symmetric array
layout and increasing the sensitivity of the observatory by 30$\%$
(Perkins et al. 2009). In its current configuration, VERITAS is able
to detect a source with a flux of 1$\%$ of the steady Crab Nebula flux
in under 30 hours. VERITAS has the capability to detect and measure
gamma rays in the 100 GeV to 30 TeV energy regime with an energy
resolution of 15-20$\%$ and an angular resolution of 0.1$^{\circ}$
(68\% containment) on an event-by-event basis.

The VHE observations presented here were made with VERITAS between
October 2008 and December 2010, sampling eight separate orbital cycles
of the binary system. The data comprise 64.5 live-time hours of
observations taken with both the original (33.3 hours) and current (31.2 hours)
array configurations. Data were taken using the complete
four-telescope array; Cherenkov images which trigger two or more
telescopes initiate a read-out of the 500 MSpS
Flash-ADC data acquisition on each PMT. The observations were made in
``wobble'' mode in which the source is offset from the center of the
field-of-view of the cameras to maximize efficiency in obtaining both
source and background measurements (Fomin et al. 1994). After
eliminating data taken under variable or poor sky conditions and
applying image cleaning methods to reject images without sufficient
signal for reconstruction, gamma-ray selection cuts were applied to
the data. These selection cuts are based on the image morphology (\textit{Mean
Scaled Width} and \textit{Length}), and the angular distance between the
reconstructed position of the source in the camera plane and the
$\textit{a priori}$ known source location. Cuts were
optimized on simulated data for both the old and new array
configurations; the analysis presented here was performed with cuts
optimized for the detection of a moderately weak (5$\%$ Crab Nebula flux) source. The data on
LS~I+61$^{\circ}$303 were taken over a relatively small range of
elevation angles (55$^{\circ}$-60$^{\circ}$) under dark skies
(little or no moonlight) resulting in an energy
threshold for these observations of 300 GeV.

In the entire integrated dataset of 64.5 hours, 176 excess events above
background were detected, corresponding to a 3.3$\sigma$ statistical
excess for steady emission, which does not constitute a significant
detection. In Tables 1-3 and Figures 1-3  the results of the three years' observations are
shown. For observations (both nightly, and binned by orbital phase)
which did not exhibit a pre-trials excess above 3$\sigma$, 99$\%$
confidence level flux upper limits are calculated using the Helene
method (Helene 1983). For nightly observations, these limits are
typically on the order of 3-10 $\times$10$^{-12}$ photons
cm$^{-2}$s$^{-1}$ above 300 GeV or 3-8$\%$ of the steady Crab Nebula
flux. When binned by orbital phase, the limits extend down to 2-6
$\times$10$^{-12}$ photons cm$^{-2}$s$^{-1}$ above 300 GeV or
2-5$\%$ of the Crab Nebula. Particularly striking is the distinct lack
of strong TeV emission during the apastron phases of $\phi=$0.5-0.8 as
these are the orbital phases during which both MAGIC and VERITAS have
previously detected strong (10-20$\%$ Crab Nebula) emission.

Unexpectedly, in late 2010, gamma-ray emission from the source was
detected over a region of the binary orbit previously undetected by
TeV instruments (Ong, 2010). From September 17 to November 7 2010 (MJD 55455-55507), a total of 13.9 hours on LS I +61$^{\circ}$ 303 were accumulated resulting in the detection of 129 excess events above background. This constitutes a detection at the 5.6$\sigma$ post-trials significance level (5.7$\sigma$ pre-trials significance with two trials for two sets of analysis cuts: standard analysis cuts and an analysis tailored for harder spectrum sources).
During phases 0.0-0.1 (close to superior conjunction at phase 0.081) the
source was observed for a total of 4.2 hours. During this time a total
of 66 excess events were recorded, corresponding to a 4.8$\sigma$ 
post-trials (5.4$\sigma$ pre-trials) significance (accounting for twenty trials accumulated by analyzing the data in ten orbital phase bins with two sets of analysis cuts). Since this detection consists of too few events to construct a differential energy spectrum with comparable statistics to those previously published, we assume a spectral index of $\Gamma = $2.6 (Acciari et al. 2008) in order to produce absolute fluxes. The source flux was measured as 6.86 $\pm$
1.45 $\times$10$^{-12}$ photons cm$^{-2}$s$^{-1}$ above 300 GeV, or
approximately 5$\%$ of the Crab Nebula flux in the same energy
range. The source presented the highest observed flux on October 8,
2010 (MJD 55477, $\phi=0.07$) with 57 excess events detected in 2.8 hours of
observations, corresponding to a 5.2$\sigma$ post-trials (5.7$\sigma$ pre-trials) significance, accounting for twenty trials accumulated by analyzing ten individual nights with two sets of analysis cuts to search for the maximal flux. During these observations the source flux was measured as
9.23$\pm$1.9 $\times$10$^{-12}$ photons cm$^{-2}$s$^{-1}$ above 300
GeV or approximately 7$\%$ of the Crab Nebula flux.

It should be noted that after appropriate statistical trials are taken
into account (for binning on nightly and orbital timescales) the
detection during phase 0.0-0.1 and on 10/08/10 stand as the only
statistically significant detections published by any TeV instrument
since 2007. The fact that these detections were made over a region
of the orbit not previously suspected to be an active TeV phase is of
particular interest.

\section{X-ray Observations}

The RXTE-PCA (Swank 1994) and \textit{Swift}-XRT (see Gehrels et
al. (2004) and Burrows et al. (2005)) data presented here were both
obtained from the public data archive at
\texttt{http://heasarc.gsfc.nasa.gov/} and analyzed using the HEASOFT (v6.9)
package (Blackburn 1995). For RXTE-PCA analysis, standard data quality
selection and screening (using the ``rex'' script in FTOOLS) was
performed. All available data were extracted using only the top layer
of Proportional Counting Unit $\#2$ in each observation. The XSPEC
12.6 software package (Arnaud 1996) was used to fit spectra with a
simple absorbed power-law, assuming a fixed absorbing hydrogen column
density (N$_{H}$) of 0.75$\times10^{22}$ cm$^{-2}$ (Kalberla et
al. 2005) between 3-10 keV and to extract fluxes in the 2-10 keV
range. Fluxes were then deabsorbed using the Wisconsin photoelectric
cross sections (Morrison and McCammon 1983), or the ``wabs'' model
within XSPEC. Also, given that the PCA instrument on RXTE is known to
give larger fluxes (on the order of 20$\%$) on the same source as other
more precise instruments (Tomsick et al. 1999), all the PCA fluxes
quoted here have been reduced by a factor of 1.18 in order to provide
a more accurate comparison to the \textit{Swift}-XRT analysis.

For the \textit{Swift}-XRT analysis, all data shown here were
accumulated in ``photon counting mode'' and analyzed using the most
recent \textit{Swift}-XRT analysis tools with the HEASOFT 6.2
package. The data were screened with standard filtering criteria and
reduced using the $\textit{xrtpipeline}$ task. Background and source
spectra were accumulated from each individual observation using
circular regions of radius 60'' and 20'', respectively. XSPEC 12.6 was
used to fit accrued spectra in the 0.3-10 keV range using a simple,
absorbed power law with the galactic hydrogen column density fixed at
0.75$\times10^{22}$ cm$^{-2}$. Flux values and associated 1~$\sigma$
statistical errors were then calculated by integrating the fitted
spectra over the 2-10 keV range.

Flux values as seen by both instruments (Figures 1-3) show a high
degree of variability with values of 2-25 $\times$10$^{-12}$
ergs cm$^{-2}$ s$^{-1}$, consistent with other observations of LS I
+61$^{\circ}$ 303 with RXTE-PCA and \textit{Swift}-XRT (i.e. Acciari et
al. 2009). 

Since variability of the X-ray emission from LS~I~+61$^{\circ}$~303
has been demonstrated on various timescales, we select only data in
the two bands which were directly overlapping in time, so as to
provide an accurate test for any correlation with VHE emission. Additionally, all of the
simultaneous X-ray observations were first checked to confirm that
there was no evidence for X-ray variability within the
observation. There are twenty simultaneous observations taken in both
X-ray and TeV (grey shaded regions) in the datasets examined here;
eight from RXTE-PCA observations and twelve from the
\textit{Swift}-XRT observations. As can be seen in Figure 4, while the
X-ray flux varies significantly, none of the points correspond to a
significant TeV detection. A test for correlation between the two
datasets results in a Pearson product-moment correlation coefficient
of -0.03$\pm$0.23; which is consistent with two uncorrelated datasets.

We note that the TeV exposures are typically significantly longer than
the corresponding X-ray observations. This is unavoidable, given the
requirement that the TeV exposure be sufficient to achieve a useful
sensitivity; however, it introduces the possibility that the
(unmeasured) X-ray flux may have varied considerably within the TeV
observation.

To test whether the correlation between X-ray and TeV emission
presented in Anderhub et al. (2009) would have been measurable in the
data presented here, we compared the Anderhub correlation fit between
X-ray and TeV data points to our data. This fit, along with the Anderhub data points from which it was derived, are shown in Figure 4. Since the energy range of our
X-ray observations was different than that presented in Anderhub et
al. (2-10 keV versus 0.3-10 keV), we have converted the Anderhub fit
to our energy range. This modification was derived by scaling the
0.3-10 keV flux range along a spectral fit with index of $\Gamma$=1.6
(consistent with spectral measurements of the source found in
Anderhub et al. 2009). Using this method, we arrive at a line fit of:
F(2-10 keV) [ 10$^{-12}$ ergs cm$^{-2}$ s$^{-1}$] = (7.8$\pm0.6$) +
(0.45$\pm0.1$)$\times$ N(E$>$300 GeV) [10$^{-12}$ cm$^{-2}$
s$^{-1}$]. According to this relation, the highest observed X-ray flux
of 14.7$\pm$0.7 $\times$10$^{-12}$ ergs cm$^{-2}$ s$^{-1}$ leads to an
expected TeV flux above 300 GeV of
15.3$^{+7.1}_{-5.2}$$\times$10$^{-12}$ photons cm$^{-2}$s$^{-1}$ or
approximately 10$\%$ of the Crab Nebula flux. For the overlapping
observation in question, VERITAS observed the source for a total of 56
minutes and measured a negative fluctuation of events relative to background corresponding to a $99\%$ Helene flux upper limit of 3.3
$\times$10$^{-12}$ photons cm$^{-2}$s$^{-1}$, a factor of 4.6 lower than the predicted value.

We note that this flux prediction does not take into account the
uncertainty in what VERITAS would have actually measured in the
observation time quoted. However, even under conservative estimates of
VERITAS sensitivity (5$\sigma$ detection of a 5$\%$ Crab-type source
in $\sim$1 hour), a source on the level predicted by the Anderhub
correlation used here would likely have yielded strong evidence for a
signal detection in the given observation livetime.

\section{Contemporaneous \textit{Fermi}-LAT GeV Observations}

We performed aperture photometry on the \textit{Fermi}-LAT dataset
 covering LS I +61$^{\circ}$ 303 for the periods during which VERITAS
 accrued observations from 2008-2010. Using the publicly available
 Fermi analysis tools and an aperture radius of 2.4$^{\circ}$ (found to
 be optimal for this source in Abdo et al. 2009), we extracted the 100
 MeV-300~GeV counts from the source in 26.5 day bins, and we used the
 $\textit{gtexposure}$ tools to calculate the exposure for each
 bin. Figure 5 shows the result of this analysis (without any
 background subtraction performed) along with the time periods during
 which VERITAS observed the source highlighted in grey regions. During
 each of the VERITAS observation windows indicated in the figure,
 VERITAS observed the source for typically 1-2 hours per night. The periods during which VERITAS successfully detected LS I +61$^{\circ}$ 303 in late 2010 (last  three shaded regions in Figure 5) do not correspond to significantly higher or lower GeV flux than the previously observed regions during which VERITAS did not detect the source.

\section{Summary and Discussion}

We have presented the results of TeV observations of the Galactic
binary LS~I~+61$^{\circ}$~303 made with the VERITAS array from October
2008 until December 2010. These observations covered eight separate
26.5 day orbital cycles with coverage (though not uniform) of all
orbital cycles. A comparison with previous observations of this source is useful. In Figure 6 all three years of observations presented in this paper, along with the previously obtained detections by both VERITAS and MAGIC (Acciari et al. 2008, Acciari et al. 2009, Albert et al. 2008) are shown. In this figure we plot only points from previous observations that had detections above 3$\sigma$ along with the currently obtained flux upper limits and detection near periastron. The previously published fluxes quoted at a different energy threshold were converted to $>$300 GeV fluxes using a spectral index of -2.5, consistent with published spectra of the source (Acciari et al. 2009). The data shown here do not indicate any significant
evidence for TeV emission during the orbital phases previously
detected by VERITAS and MAGIC (i.e., near apastron passage), with flux upper limits derived which are $\sim$2-3 times lower than the previously detected emission. Instead,
the source has been detected for the first time during orbital phases
0.0-0.1, close to superior conjunction (at $\phi=0.081$). A further
implication of these observations is that there is no direct evidence
that the process responsible for the detected $\textit{Fermi}$-LAT emission continues beyond the published 6~GeV cut-off: all published spectra show only non-contemporaneous
data.

An important caveat to these points is that we cannot rule out the
existence of near-apastron TeV emission during the 2008-2010
timeframe. As is the case with all TeV observations, the sampling of
the VHE flux from LS I +61$^{\circ}$ 303 was unavoidably limited, with
observations taking place separated by days and only for integrations
lasting several hours. Furthermore, since the orbital period of LS I +61$^{\circ}$ 303 ($\sim$27 days) is very close to the lunar cycle; full-orbit observations of LS I +61$^{\circ}$ 303 with ground-based TeV telescopes are limited by their inability to observe during bright moon phases. The observations sampled only 8 orbits out of 27 covered by these two years,
and observations in 2009-2010 focussed exclusively on orbital phases
$\phi=0.5-0.8$. This leaves open the possibility that the source was
active at VHE in some of the unobserved orbits during apastron
passage. Also, the possibility exists that the sampling provided by
the VERITAS observations simply missed the significant TeV activity.

We have also presented both contemporaneous and directly overlapping
X-ray observations taken with \textit{Swift}-XRT and RXTE-PCA. The 20
directly overlapping X-ray observations do not show any evidence for
correlation with the TeV emission. This stands in contrast somewhat to
the result of Anderhub et al. (2009), in which a correlation was
detected between X-ray and TeV emission. However, the result of
Anderhub et al. (2009) was obtained during a TeV outburst covering 60\%
of a single orbit, and this paper makes no claim of longer-term correlation
between the two emission bands. We see no evidence for such a
correlation during the observations taken here, which were obtained
over several different orbital cycles.

During the 2008-2010 period, the GeV flux was monitored by
\textit{Fermi}-LAT. Dubois et al. (2010) have shown that the most
recent \textit{Fermi}-LAT GeV observations showed a marked flux
increase, of $\sim$40$\%$, in March 2009, after which the previously
clear orbital GeV flux modulation is no longer measurable. The result
of Torres et al. (2010b) shows that, while the X-ray periodicity of the
source is well defined, it is not entirely stable; the degree of
modulation, and the phase of the peak of the emission varies from
orbit-to-orbit and year-to-year. These results, compounded by the TeV
measurements in this paper, indicate that the non-thermal emission
from LS~I~+61$^{\circ}$~303 is not as stable as its southern hemisphere
counterpart, LS~5039.

Torres (2010a) has suggested that the GeV emission characteristics can
be simply explained by invoking a two-component model, where the
observed flux is the result of magnetospheric (pulsar) emission, plus
a wind contribution produced either in the inter-wind region and/or
the pulsar wind zone. The magnetospheric contribution would be steady,
and exhibit a sharp spectral cut-off at $\sim6$~GeV, while the wind
contribution would be expected to vary naturally with orbital
phase. If the rate of particle acceleration is the main factor
determining the GeV and TeV fluxes, then the observations reported here
do not seem to support this idea in its simplest form. Since the wind
component is expected to be sensitive to orbital modulation, and
it provides the primary contribution to the TeV emission, the GeV flux
increase since March 2009 should naturally lead to a larger fraction
of the flux being modulated with the orbital period, and to increased
TeV flux. The opposite of both of these effects is observed. Other
factors may affect this conclusion, however; for example, increased
inverse-Compton cooling may cause the spectrum to steepen, enhancing
the GeV flux and decreasing the TeV flux. The system may also have moved
into a state where the Compton process has saturated, and the output
spectrum is largely independent of the input photon flux, thus
explaining the disappearance of orbital modulation.

How can this apparent long term variability be explained? The results
of radio monitoring (Gregory 2002) show a definite modulation of the
phase and peak flux density of the radio outbursts with a period of
$1667\pm8$ days. In Gregory and Neish (2002) this effect is attributed
to a model where a pulsar is embedded in a co-planar fashion within
the equatorial disk of the Be star. The long term modulation then
results from periodic enhancements of the stellar wind density within
the disk. If this is the case, and the HE-VHE emission results from
the same particle population as the radio emission, then perhaps the
long term variation in these bands is not unexpected. It certainly
seems likely, given the contrast with the stability of the orbital
lightcurves of LS~5039 (Kishishita et al. 2009, Aharonian et al. 2006,
Abdo et al. 2009), that variations in the circumstellar environment of
the Be star play an important role in determining the variability of
the observed emission. Sensitive TeV observations of
LS~I~+61$^{\circ}$~303 began only in September 2005, while
\textit{Fermi} was launched in June 2008. Continued monitoring over
the coming years from X-ray to TeV will allow us to probe longer-term
cycles, to study the cross-correlations in more detail and to
characterize the emission spectra more precisely.

\acknowledgments

This research is supported by grants from the U.S. Department of
Energy, the U.S. National Science Foundation, the Smithsonian
Institution, by NSERC in Canada, by Science Foundation Ireland (SFI 10/RFP/AST2748) and by
STFC in the U.K..  We acknowledge the excellent work of the technical
support staff at the Fred Lawrence Whipple Observatory and the collaborating institutions in the
construction and operation of the instrument.

The submitted manuscript has been created by employees of UChicago Argonne, LLC,
Operator of Argonne National Laboratory (``Argonne'') in conjunction with the VERITAS collaboration. Argonne, a U.S.
Department of Energy Office of Science laboratory, is operated under
Contract No. DE-AC02-06CH11357. The U.S. Government retains for
itself, and others acting on its behalf, a paid-up nonexclusive,
irrevocable worldwide license in said article to reproduce, prepare
derivative works, distribute copies to the public, and perform
publicly and display publicly, by or on behalf of the Government.

J. Holder acknowledges the support of the NASA Fermi Cycle 2 Guest
Investigator Program (grant number NNX09AR91G).

\begin{table}

\begin{center}

 \begin{tabular}{c|c|c|c|c}

\textbf{MJD}&  \textbf{Phase} & \textbf{Livetime}  & \textbf{Significance}  & \textbf{Flux/99$\%$ U.L.$ >$ 300 GeV}\\ 
\textbf{(UTC)}&\textbf{($\phi$)}&\textbf{min}&$\sigma$&\textbf{10$^{-12}$ $\gamma$ $ $cm$^{-2}$ s$^{-1}$}\\\hline
 54760.3 & 0.03 & 113.6 & 2.2 &$<$10.1   \\\hline
 54761.4 & 0.07 & 89.6  & 1.7 &$<$9.9  \\\hline
 54762.3 & 0.1  & 90.2  &  0.6 & $<$6.8  \\\hline
 54763.4  & 0.15 & 18.1  & -0.4& $<$10.3  \\\hline
 54764.2 & 0.18 & 111.5 & 0.5 & $<$4.4  \\\hline
 54765.2 & 0.22 & 126.9  & -1.4& $<$2.6  \\\hline
 54766.2 & 0.26 & 90.4  & 0.2 & $<$5.9  \\\hline
 54767.3 & 0.31 & 36.1  & 0.2 & $<$6.9  \\\hline
 54768.3 & 0.35 & 90.7  & 0.4 & $<$6.0  \\\hline
 54769.3  & 0.37 & 109.7 & -0.3& $<$2.9  \\\hline
 54770.3 & 0.41 & 54.7  & -0.9& $<$4.4  \\\hline
 54771.3 & 0.44 & 72.8  & 1.0 & $<$6.5  \\\hline
 54772.3 & 0.48 & 72.9  & 1.7 & $<$10.2 \\\hline
 54774.3 & 0.56 & 90.7  & 2.7 & $<$9.2  \\\hline
 54775.3 & 0.59 & 54.1  & -0.7& $<$4.4   \\\hline
 54776.3 & 0.63 & 162.5 & 1.5 & $<$4.4  \\\hline
 54777.4 & 0.67 & 72.5   & 0.5 & $<$7.2  \\\hline
 54778.4 & 0.71 & 54.4   & -0.8& $<$5.8  \\\hline
 54779.4 & 0.75 & 36.5  & -0.3& $<$7.8  \\\hline
 54856.1  & 0.64 & 102.9 & 0.4 & $<$6.1  \\\hline
 54857.1 & 0.68 & 90.5  & -1.8& $<$2.7  \\\hline
 54861.1 & 0.83 & 95.2   & 1.1 & $<$7.4   \\\hline
 54862.1  & 0.87 & 104.0 & 0.9 & $<$6.8  \\\hline
 54863.1  & 0.91 & 76.1  & 1.4 & $<$8.8   \\\hline

 \end{tabular}
\end{center}
\caption{VERITAS observations of LS I +61$^{\circ}$~303 in the 2008-2009 observing season.}
\end{table}

\begin{table}
 \centering

 \begin{tabular}{c|c|c|c|c}

\textbf{MJD}&  \textbf{Phase} & \textbf{Livetime}  & \textbf{Significance}  & \textbf{Flux/99$\%$ U.L.$ >$ 300 GeV}\\ 
\textbf{(UTC)}&\textbf{($\phi$)}&\textbf{min}&$\sigma$&\textbf{10$^{-12}$ $\gamma$ $ $cm$^{-2}$ s$^{-1}$}\\\hline
55119.3 & 0.58 & 128.5  & 1.6  & $<$7.3\\\hline
55120.3 & 0.62 & 107.1  & -0.4 & $<$5.5\\\hline
55121.3  &0.65  &161.6  & -2.5 & $<$1.6\\\hline
55122.3 &0.69 & 107.9  & 0.9 & $<$6.0\\\hline
55123.3 &0.73  &18.2   &  -0.7 &$<$9.4\\\hline
55124.3 &0.77 & 99.2  &  1.9 & $<$7.9\\\hline
55146.2 &0.59 & 146.3 &  3.4 & 5.8$\pm$1.9 \\\hline
55151.2 &0.78 & 123.2  &  2.1 & $<$10.0\\\hline
55175.2 &0.69 & 54.7 &   -0.7& $<$6.5\\\hline
55176.2 &0.73 & 54.6   &  -0.3& $<$7.3\\\hline
55177.2 &0.76 & 18.2  &   -0.8& $<$10.3\\\hline

\end{tabular}

\caption{The same as Table 1, but for the 2009-2010 observing season.}
\end{table}

\begin{table}
 \centering

 \begin{tabular}{c|c|c|c|c}

\textbf{MJD}&  \textbf{Phase} & \textbf{Livetime}  & \textbf{Significance}  & \textbf{Flux/99$\%$ U.L.$ >$ 300 GeV}\\ 
\textbf{(UTC)}&\textbf{($\phi$)}&\textbf{min}&$\sigma$&\textbf{10$^{-12}$ $\gamma$ $ $cm$^{-2}$ s$^{-1}$}\\\hline
55455.4 &   0.26 &   55.4 &     1.7  &     $<$10.08   	 \\\hline
55457.4 &   0.33 &  18.0  &    1.4   &     $<$18.00      \\\hline
55476.4 &  0.05  &  91.7  &    1.3   &     $<$11.4       \\\hline
55477.3 &   0.09 &  162.6 &    5.7   &    9.23 $\pm$ 1.9 \\\hline
55480.3 &  0.2   &  54.2  &    2.3   &     $<$19.7       \\\hline
55481.3 &   0.24 &  107.9 &     2.0 &     $<$9.3        \\\hline
55482.3 &   0.28 &  108.3 &     -0.3 &     $<$4.2        \\\hline
55483.3 &  0.32  &  128.1 &   -0.04  &     $<$5.84       \\\hline
55505.3 &  0.14  &  36.5  &   2.7    &     $<$23.7       \\\hline
55506.2 &  0.18  &  73.3  &  0.9     &     $<$11.09   	 \\\hline

\end{tabular}

\caption{The same as Table 1, but for the 2010-2011 observing season.}
\end{table}

\begin{figure}
\begin{center}
   \includegraphics[width=\textwidth,height=110mm]{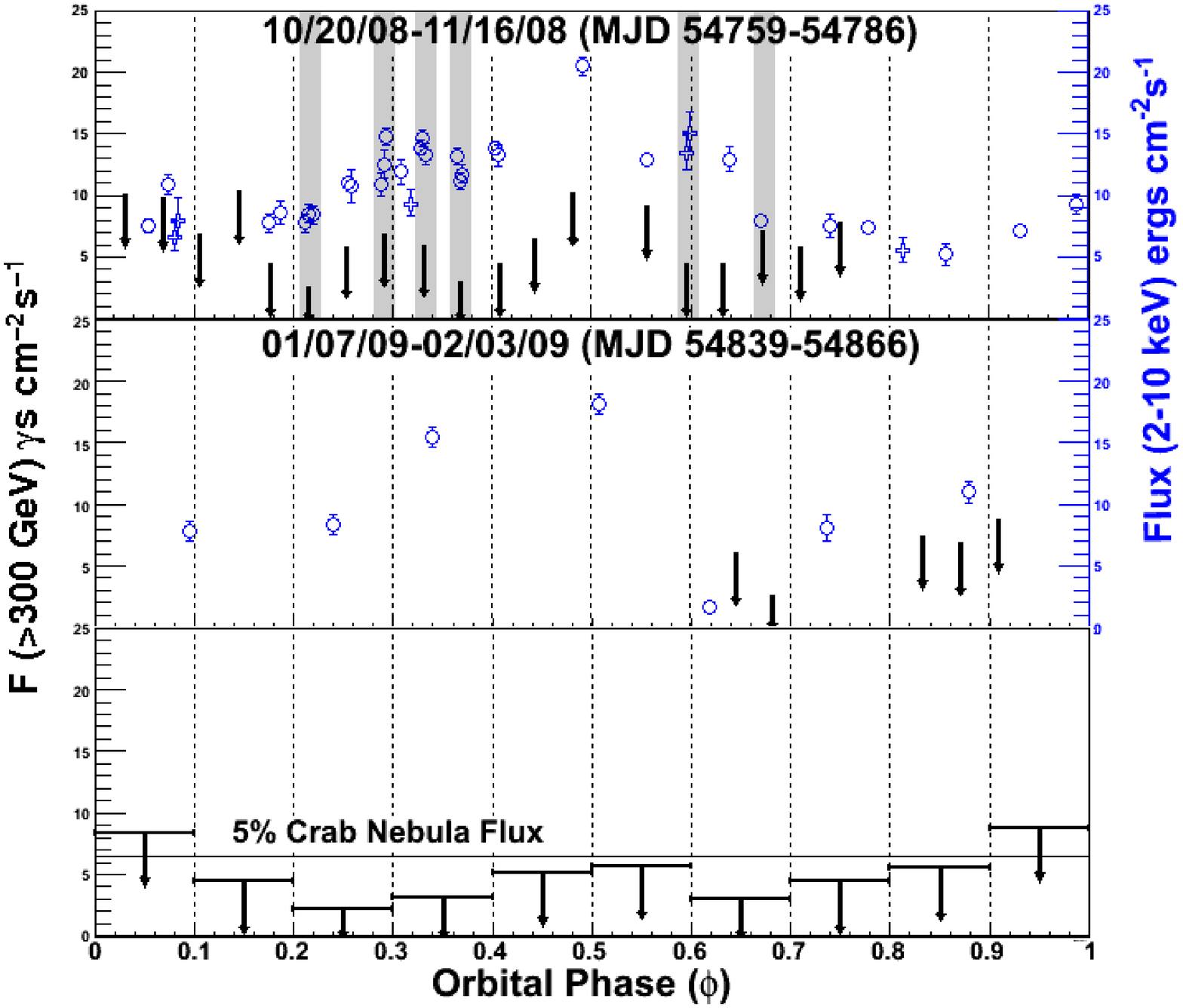}
\end{center}
\caption[]{TeV flux upper limits from VERITAS (black points and arrows) compared to contemporaneous X-ray fluxes as measured by \textit{Swift}-XRT (blue crosses) and RXTE-PCA (blue open circles). The bottom panel shows the integrated result from the two measured orbital cycles observed from October 2008 to February 2009. VERITAS flux measurements with less than 3$\sigma$ pre-trials significance are shown as 99$\%$ confidence flux upper limits. The shaded regions indicate observations which had directly overlapping X-ray measurements by either RXTE or \textit{Swift}. }
\end{figure}

\begin{figure}
\begin{center}
   \includegraphics[width=\textwidth,height=150mm]{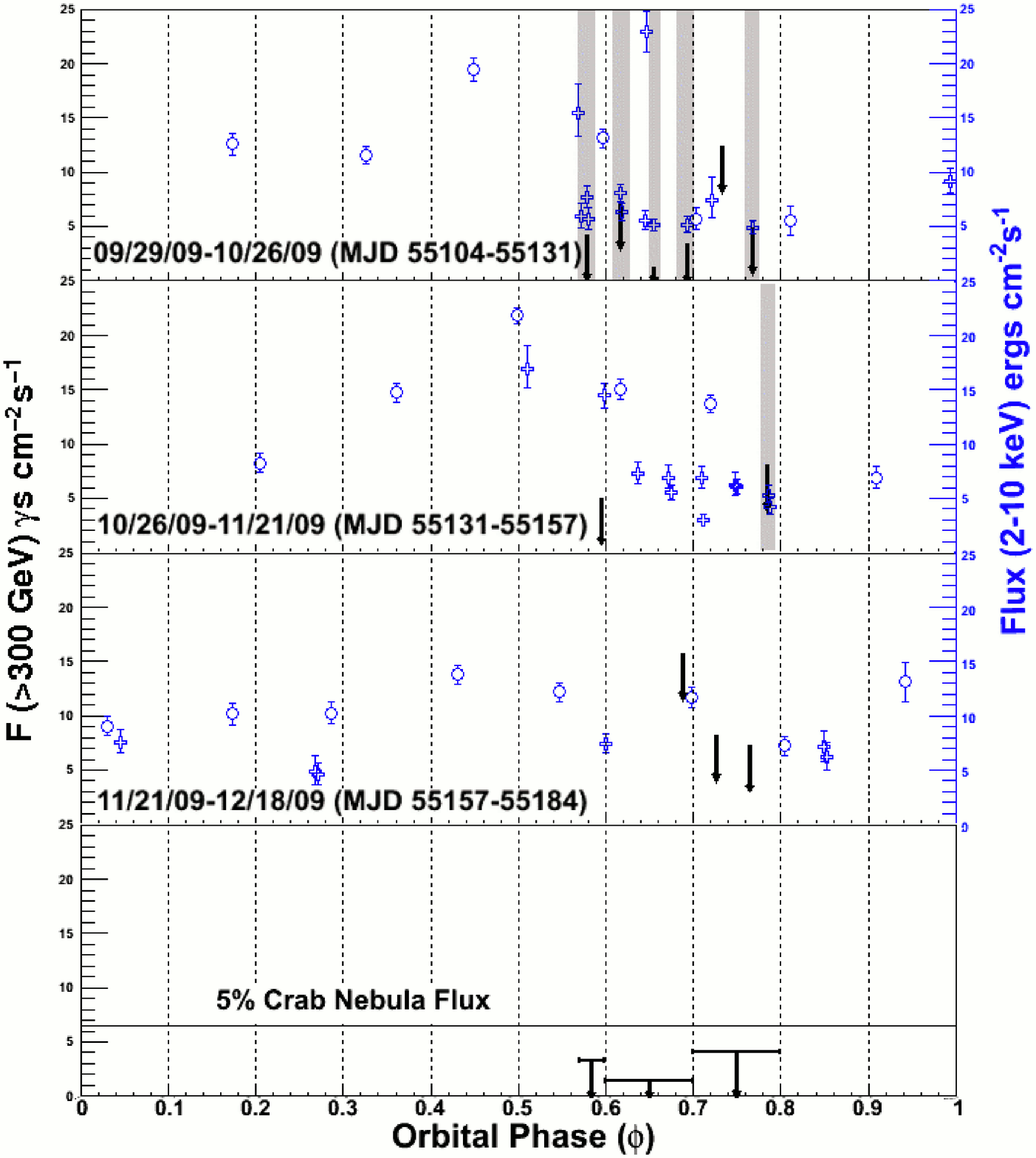}
\end{center}
\caption[]{The same as Figure 1, but for the 2009-2010 observing season.}
\end{figure}

\begin{figure}
\begin{center}
   \includegraphics[width=\textwidth,height=150mm]{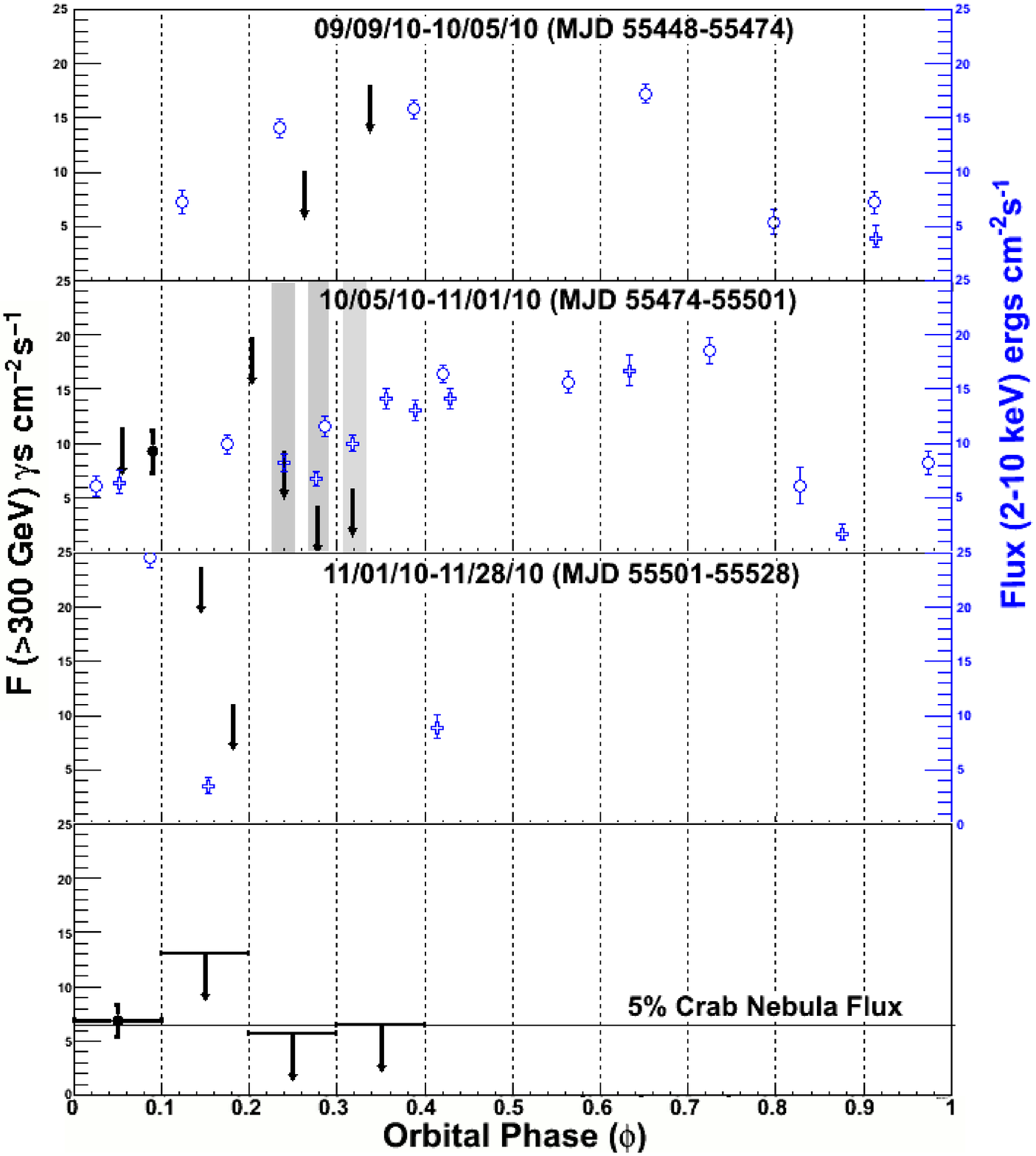}
\end{center}
\caption[]{The same as Figure 1, but for the 2010-2011 observing season.}
\end{figure}

\begin{figure}
\begin{center}
   \includegraphics[width=120mm,height=85mm]{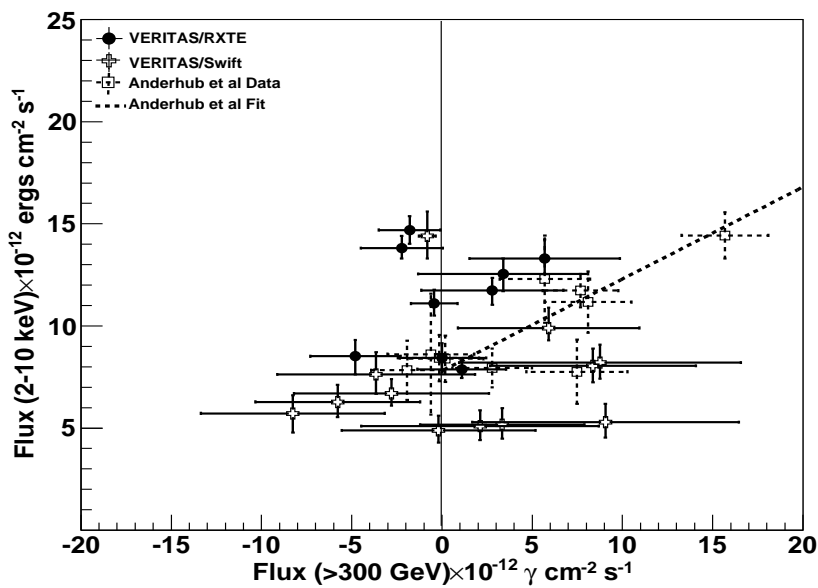}
\end{center}
\caption[]{A comparison of the strictly simultaneously observed TeV and X-ray fluxes seen by VERITAS and $\textit{Swift}$/RXTE, along with the data points and associated correlation fit from Anderhub et al. (2009).}
\end{figure}

\begin{figure}
\begin{center}
   \includegraphics[width=120mm,height=70mm]{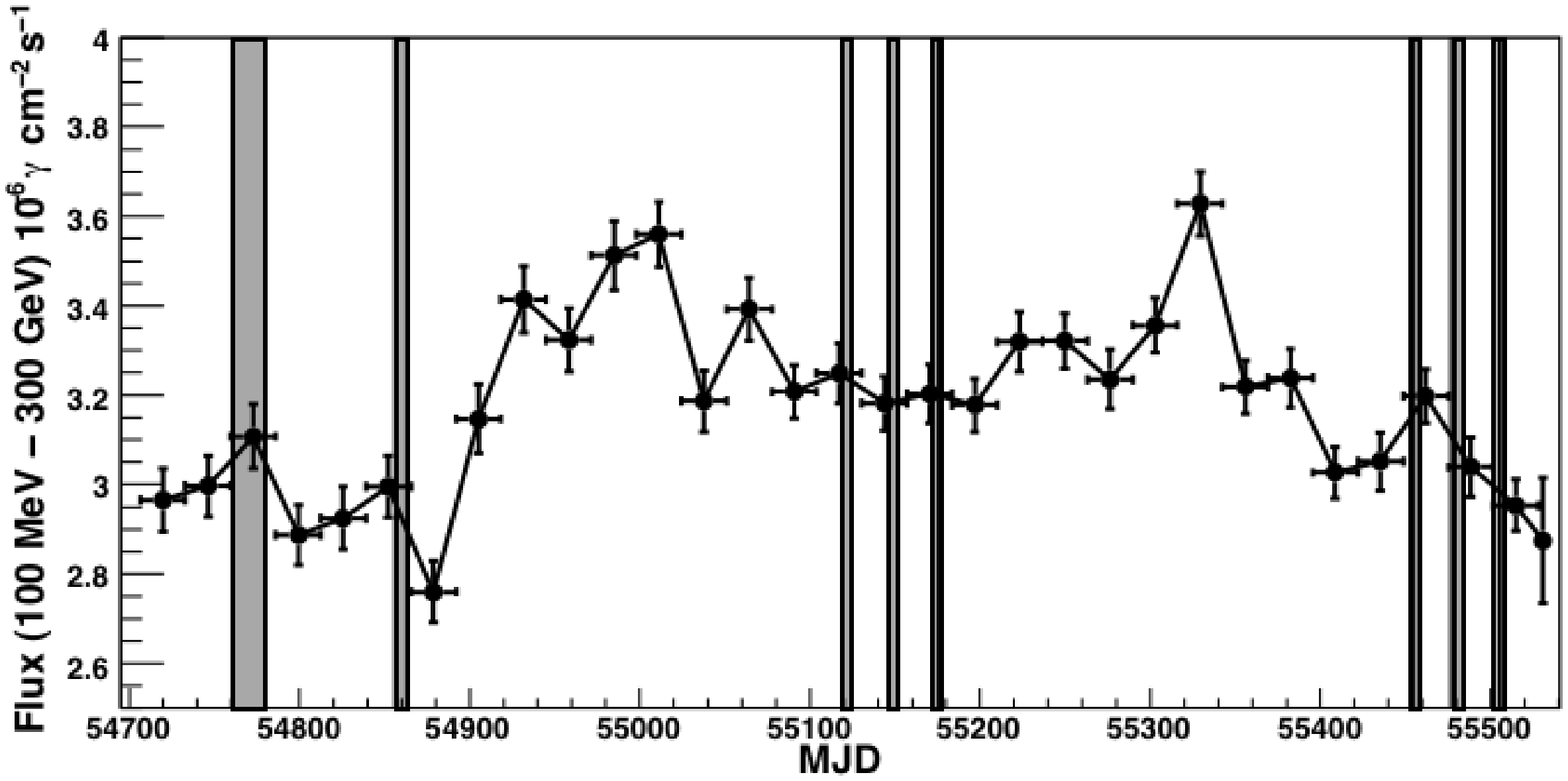}
\end{center}
\caption[]{The GeV flux observed by \textit{Fermi}-LAT from LS I +61$^{\circ}$ 303 (with no background subtraction performed). The grey shaded regions show the times during which VERITAS observed the source during 2008-2010.}
\end{figure}

\begin{figure}
\begin{center}
   \includegraphics[width=120mm,height=85mm]{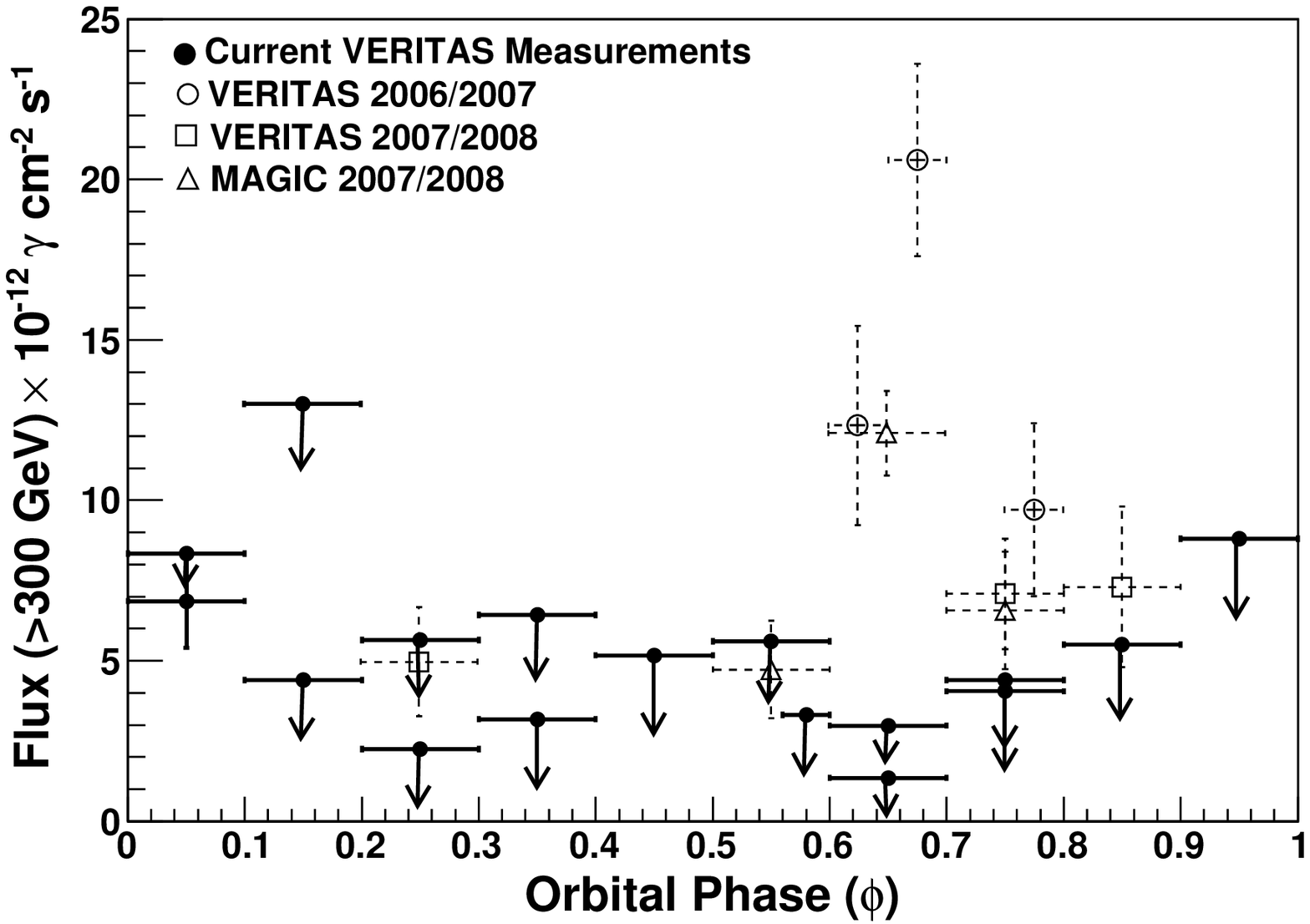}
\end{center}
\caption[]{The observations presented in this paper (solid lines) along with previous observations (dashed lines) from Acciari et al (2008), Acciari et al (2009), and Albert et al. (2008). Only detections at a significance larger than 3$\sigma$ are shown from previous observations.}
\end{figure}


\begin{references}

\reference{1} Abdo, A.A. et al., 2009, ApJ, 701:L123-L128

\reference{2} Abdo, A.A. et al., 2010, ApJSupp., 188, 405A

\reference{3} Acciari, V. et al. 2008, ApJ, 679, 1427

\reference{4} Acciari, V. et al. 2009, ApJ, 700, 2

\reference{5}Aharonian, F.A. et al., 2006, A\&A, 460, 743

\reference{6}Albert, J. et al. 2006, Science, 312, 1771

\reference{7}Albert, J. et al. 2008, 684, 1351

\reference{8}Anderhub, H., et al. 2009, ApJ, 706, L27

\reference{9}Aragona, C.  et al., 2009 ApJ 698, 514,

\reference{10} Araudo, A.T. 2009, A$\&$A 503, 673

\reference{11} Arnaud, K.A. 1996, Astronomical Data Analysis Software and Systems V, eds. G. Jacoby, and J. Barnes, ASP Conf. Series, volume 101, p17

\reference{12} Blackburn, J.K. 1995, Astronomical Data Analysis Software and Systems IV, eds. R.A. Shaw, H.E. Payne, and J.J. E. Hayes, ASP Conf Ser. ,Vol 77, p367

\reference{13} Burrows, D.N. et al. 2005, Space Sci. Rev., 120, 165

\reference{14} Casares, J. et al. 2005, MNRAS, 360, 1105.

\reference{15} Dhawan, V. et al. 2006, in Proc. of Microquasars and Beyond: From Binaries  to Galaxies, in Proceedings of Science, Como, IT. ed. Belloni, T , p 52

\reference{16} Dubus, G., Cerutti, B., and Henri, G., 2010,  A$\&$A 516, A18

\reference{17} Dubois, R. et al., 2010,  in Proc. ``1st Sant Cugat Forum on Astrophysics; ICREA Workshop on the high-energy emission from pulsars and their systems'', Sant Cugat, Spain. April 2010

\reference{18} Fomin, V.P. et al. 1994, Astroparticle Physics, 2, 137

\reference{19} Gehrels, N. et al., 2004, ApJ, 611, 1005

\reference{20} Gregory, P.C. 2002, ApJ, 525, 427

\reference{21} Gregory, P. C. and Neish C. 2002, ApJ, 580, 1133

\reference{22} Greiner, J. and Rau, A. 2001, A$\&$A, 375, 145

\reference{23} Harrison, F.A. et al. 2000, ApJ, 528, 454

\reference{24} Helene, O. 1983, Nuc. Inst. and Meth., 212, 319

\reference{25} Hermsen, W. et al. 1977, Nature, 269, 494

\reference{26} Hutchings, J.D. and Crampton, D. 1981, PASP, 93, 486

\reference{27} Kalberla, P. M. W. et al. 2005, A$\&$A, 440, 775

\reference{28} Kishishita, T. et al. 2009, ApJ, 697, L1

\reference{29} Massi, M. et al. 2001, A$\&$A, 376, 217

\reference{30} Massi, M. and  Zimmerman, L.,  2010, A$\&$A, 515, 82

\reference{31} Morrison, R. and McCammon, D., 1983,  ApJ 270, 119

\reference{32} Ong, R.A.  et al., 2009, in “Proc 31st ICRC”, Lodz, Poland, arXiv:0912.5355 

\reference{33} Ong, R.A. et al., 2010, Astronomers Telegram $\#$2948

\reference{34} Perkins, J. A. et al. 2009 Fermi Symposium, eConf Proceedings C091122

\reference{35}  Romero, G.E. et al. 2007, A$\&$A 474, 15–22 

\reference{36} Sierpowska-Bartosik, A. and Torres, D., 2009, ApJ, 693, 1462

\reference{37} Smith, A., Falcone, A., Holder, J., Kaaret, P., Maier, G., and Pandel, D. 2009, ApJ, 693, 1621

\reference{38}  Swank, J.H. 1994, in Proceedings of 181st American Astronomical Society Meeting, Phoenix, AZ, 185, 6701

\reference{39} Tavani, M. et al., 1998, ApJ, 497, L89

\reference{40} Tomsick, J. A., Kaaret, P., Kroeger, R. A.,  Remillard, R.A., 1999, ApJ, 512, 892

\reference{41} Torres, D. et al., 2010a, in Proc. ``1st Sant Cugat Forum on Astrophysics; ICREA Workshop on the high-energy emission from pulsars and their systems'', Sant Cugat, Spain, April, 2010, arXiv:1008.0483

\reference{42} Torres, D. et al., 2010b, ApJ, 719, L104

\end{references}
\end{document}